\documentclass[twocolumn,showpacs,superscriptaddress,prb,floatfix]{revtex4}
\usepackage{graphicx} \usepackage{epsfig} \usepackage[usenames]{color}
\usepackage{subeqn} 
\usepackage{ulem} 

\newcommand{\ket}[1]{\ensuremath{|#1\rangle}}
\newcommand{\bracket}[2]{\ensuremath{\langle #1|#2\rangle}}

\newcommand{\bk}{{\bf k}} \newcommand{\bt}{{\bf t}}
\newcommand{\bx}{{\bf x}}

\newcommand{\hot}{\hat{\bf t}}

\newcommand{\bgamma}{\mbox{\boldmath$\gamma$}}

\newcommand{\bA}{{\bf A}} \newcommand{\bB}{{\bf B}}
 \newcommand{\bS}{{\bf S}}
\newcommand{\bT}{{\bf T}}

\newcommand{\kB}{{k_{\text{B}}}} 

\date{\today}

\begin{document}
\title{Interaction of two level systems in amorphous materials with arbitrary phonon fields} 
\author{D. V. Anghel}
\affiliation{National Institute for Physics and Nuclear Engineering--''Horia Hulubei'', Str. Atomistilor no.407, P.O.BOX MG-6, Bucharest - Magurele, Romania}

\author{T. K{\"u}hn}
\affiliation{Nanoscience Center, Department of Physics, University of Jyv\"askyl\"a, P.O. Box 35, FIN-40014 University of Jyv\"askyl\"a}

\author{Y. M. Galperin}
\affiliation{Department of
Physics \& Centre of Advanced Materials and Nanotechnology, University
of Oslo, PO Box 1048 Blindern, 0316 Oslo,~Norway}
\affiliation{Argonne National Laboratory, 9700 S. Cass
Av., Argonne, IL 60439, USA}
\affiliation{ A. F. Ioffe
Physico-Technical Institute of Russian Academy of Sciences, 194021
St. Petersburg, Russia}

\author{M. Manninen}
\affiliation{Nanoscience Center, Department of Physics, University of Jyv\"askyl\"a, P.O. Box 35, FIN-40014 University of Jyv\"askyl\"a}

\begin{abstract}
To describe the interaction of the two level systems (TLSs) of an
amorphous  solid with arbitrary strain fields, we introduce a
generalization of  the standard interaction Hamiltonian. In this new
model, the interaction  strength depends on the orientation of the TLS
with respect to the strain  field through a $6\times 6$ symmetric
tensor of deformation potential  parameters, $[R]$. Taking into
account the isotropy of the amorphous  solid, we deduce that $[R]$ has
only two independent parameters.  We show how these two parameters can
be calculated from experimental data  and we prove that for any
amorphous bulk material the  average coupling of TLSs with
longitudinal phonons is always stronger  than the average coupling
with transversal phonons (in standard notations,
$\gamma_l>\gamma_t$).
\end{abstract}
\maketitle

The thermal properties of dielectric crystals at low temperatures are
well described by the Debye model. If the temperature is much smaller
than the Debye temperature of the crystal, then the optical phonon
modes  are not excited and the only contribution to the heat capacity
and  heat conductivity comes from the acoustic phonons. In three
dimensional (3D)  systems, low-frequency acoustic phonons have a
linear dispersion relation,  $\omega = c_{t,l} k$, where $\omega$ is
the angular frequency,  $c_t$ and $c_l$ are the transversal and
longitudinal sound velocities,  respectively, and $k$ is the absolute
value of the phonon's wavevector,  $\bk$. This gives a specific heat
proportional to the temperature to the power three ($c_V\propto T^3$).

  A good estimate for the heat conductivity, $\kappa$, is
  $\kappa=\frac{1}{3}c_V c l$, where $c$ is an average sound velocity
  and $l$ is the phonon mean free path, which depends not only on the
  material, but also on the sample quality. Impurities or lattice
  defects, even at low concentration, reduce the phonon mean free
  path and in this way may decrease dramatically the heat
  conductivity.
  \cite{AshcroftMermin:book,Ziman:book,Kittel:book,zellerpohl:physrev}
  As a result, the temperature dependence of the heat conductance is
  determined by the dependence of the phonon mean free path on its
  energy. Since such dependences can be very much different for
  different phonon scattering mechanisms, the resulting temperature
  dependence of $\kappa$ can be in general rather complicated.

In high-quality crystals of relatively small size and at sufficiently
  low temperature,  the phonon mean free path may become comparable to or
  bigger than the crystal dimensions. In this case the phonons will
  scatter mainly at the surfaces and the mean free path is limited by
  the surface diffusivity  and the geometrical features of the sample.
  \cite{Klitsner:PRB} In this case $l$ is independent of the phonon
  frequency and $\kappa \propto c_V \propto T^3$.

Continuing to decrease the temperature and the size of the system,  we
get into the \textit{mesoscopic} regime, where one or more dimensions
of the system become comparable to the dominant phonons wavelength.
Typically we find this regime in nanometer-size objects, at
temperatures  of a few Kelvins or less.  At such scales, the phonon
interaction with the surfaces becomes  important, since it leads to
coupling between different vibrational modes.  This can lead to a very
complicated set of phonon modes, with  nonlinear dispersion
relations. One example are the \textit{Lamb modes}\cite{Auld:book},
which are among the eigenmodes of  a free standing infinite membrane.
Due to the boom of nanotechnology, mesoscopic systems are used  in
more and more practical applications  (see for example
Refs.~\onlinecite{Rep.Prog.Phys.58.311,RevModPhys.78.217,APL.72.2250,LeivoPekola:APL72,PhysRevLett.81.2958,Leivoetal:ApplS5,PhysRevB.70.125425,PhysicaB.284.1968}).
The physical properties of mesoscopic devices can differ dramatically
from those of bulk systems, the difference being more pronounced at
lower  temperatures.

Important parts of many mesoscopic devices -- such as microbolometers,
electromechanical sensors and actuators -- are ultrathin free standing
membranes.  The thermodynamics and thermal transport of such membranes
are specifically important at low temperatures, where heat release is
a bottleneck for device performance. At the same time, the
low-temperature transport properties are  rather unusual.  For example
in many experiments the heat  conductivity, $\kappa$, of large, thin
membranes or long, narrow and thin bridges is proportional to $T^p$,
where $p$ takes values between 1.5 and 2.
\cite{RevModPhys.78.217,APL.72.2250,LeivoPekola:APL72,PhysRevLett.81.2958}
The specific heat of mesoscopic membranes is more difficult to
measure.  It can, however, be extracted from the amplitude of the
temperature  oscillations in AC measurements and it appears also to be
proportional  to $T^p$, where $p$ lies roughly between 1 and
2.\cite{LeivoPekola:APL72}

The temperature dependence of $\kappa$ and $c_V$ in mesoscopic
membranes and bridges can partly be explained by a crossover
from a three-dimensional to a two-dimensional phonon gas distribution. 
This crossover takes place 
when the dominant thermal phonon wavelength,
$\lambda_T=2\pi\hbar c/\kB T$,
is comparable to the membrane thickness.

Because of the finite thickness, there are gaps in the phonon spectra, the
phonon dispersion relations near the band edges being nonlinear. As a result,
the temperature dependence of the specific heat deviates from the law
$c_V\propto T^2$, which one would intuitively expect from a two-dimensional
phonon gas. Instead, one obtains for very low temperatures $c_V\propto T$.
\cite{LeivoPekola:APL72,PhysRevLett.81.2958,PhysRevB.59.9854,PhysRevB.70.125425}

To study the heat conductivity in mesoscopic insulators, bridges 
have been cut out of the above mentioned membranes. There, a decrease of the 
exponent of the temperature
dependence of $\kappa$ from $2$ to about $1.5$ was observed. Furthermore, the 
cut-off frequency of the temperature oscillations in AC heating measurements
($f_c\propto\kappa/c_V$) was observed to increase with temperature for the 
narrowest bridges.\cite{Leivo:thesis,Luukanen:thesis,LeivoPekola:APL72} 
Taking into account the fact that the edges of the 
bridges are very rough due to the cutting process, the measured data 
could be explained by using the dispersion relations of the above 
mentioned Lamb modes to calculate heat capacity and heat 
conductivity.\cite{PhysRevB.70.125425} 

Nevertheless, in some experiments the same behavior, $\kappa\propto T^p$ 
with $p$ being roughly 2,\cite{Leivo:thesis,Luukanen:thesis,LeivoPekola:APL72,APL.72.2250,PhysicaB.284.1968} and the increase of $f_c$ with the temperature 
\cite{Leivo:thesis,LeivoPekola:APL72} was observed also above the
2D--3D crossover  temperature. 
To explain these features we have to extend our model and 
take into account the amorphous structure of the material 
-- low stress amorphous silicon nitride -- 
used in the above mentioned experiments. 

\section{Two Level Model}

Amorphous or glassy materials 
differ significantly from crystals, especially in the low 
temperature range, where, for 3D bulks, $\kappa\propto T^2$ and
$c_V\propto T$.\cite{zellerpohl:physrev,esquinazi:book} 
These temperature dependences 
were explained by the presence of specific dynamic defects. These 
defects are modeled by an ensemble of so-called 
\textit{two level systems} (TLS) that exists in the material.\cite{philips:lowtemp,anderson:philmag} A TLS can
be understood as an atom, or a group of atoms, which can tunnel between
two close minima in configuration space, forming a hybridized doublet state.
The presence of such minima is a hallmark of the glassy state. 
If the energy splitting between these minima is $\lesssim \kB T$, then the
TLS can be excited from its ground state onto the upper level. 
In this way the TLSs contribute to the heat capacity. TLSs can also 
scatter phonons and in this way decrease
their mean free path and, correspondingly, the heat conductance.

\begin{figure}[t]
\begin{center}
\unitlength1mm\begin{picture}(80,30)(0,0)
\put(0,0){\epsfig{file=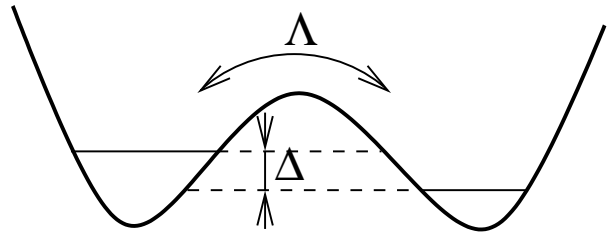,width=80mm}}
\end{picture}
\caption{An atom of a group of atoms moves in an effective potential like 
this. The separation between the ground states in the two wells, $\Delta$, 
is much smaller than the energy scale of the oscillation frequency 
in the wells, $\omega_0$.}
\label{TWP}
\end{center}
\end{figure}

An effective double-well potential and the tunneling of the 
atom between the two wells are depicted in
Fig. \ref{TWP}.\cite{philips:lowtemp,anderson:philmag} Written in the
2D Hilbert space  spanned by the ground states of the two wells, the
effective Hamiltonian of this TLS reads  
\begin{equation}
  \label{eqn_TLS_hamiltonian}
  H_{\text{TLS}}
      =  \frac{\Delta}{2}\sigma_z -\frac{\Lambda}{2}\sigma_x 
 \equiv \frac{1}{2}\left(\begin{array}{cc} 
                     \Delta   & -\Lambda\\ 
                     -\Lambda & -\Delta \end{array}
                \right) 
\end{equation}
with $\Lambda$ describing the tunneling between the two wells. 
In general, $\Delta$ is called the {\em asymmetry of the potential} 
and $\Lambda$ is called the {\em tunnel splitting}. The Hamiltonian 
(\ref{eqn_TLS_hamiltonian}) may be diagonalized by an orthogonal 
transformation $O$, 
\begin{equation}
\label{eqn_TLS_hamiltonian_diag}
 H'_{\text{TLS}} \equiv O^T H_{\text{TLS}} O 
 = \frac{\epsilon}{2}\sigma_z
 = \frac{1}{2}\left(\begin{array}{cc}
                \epsilon &0\\
                 0 &-\epsilon\end{array}\right) 
\end{equation}
where $\epsilon\equiv\sqrt{\Delta^2+\Lambda^2}$ is the excitation 
energy of this TLS and by the superscript $T$ we denote 
in general \textit{the transpose of a matrix}. Since the TLS 
can be in only two states, let us denote the ground state by 
$|\! \downarrow\rangle$ and the excited state by $|\! \uparrow\rangle$.
Phonons at frequencies close to the level splitting $\epsilon$
are strongly scattered by TLSs.

The Hamiltonian parameters $\Delta$ and $\Lambda$ are distributed 
with the density $VP(\Delta,\Lambda)$, where $V$ is the volume of the solid. 
According to the standard tunneling model (STM), 
$P(\Delta,\Lambda)$ is assumed to have the form 
\begin{equation}
P(\Delta,\Lambda) =P_0/\Lambda
\,, 
\label{eqn_distribution_delta_lambda}
\end{equation}
where $P_0$ is a constant.
If expressed through the variables $\epsilon$ and $u\equiv\Lambda/\epsilon$,
the distribution function is
\begin{equation}
P(\epsilon,u)  = \frac{P_0}{u\sqrt{1-u^2}}
\,.\label{eqn_distribution_E_u}
\end{equation}

The strain caused by a phonon or any other kind of distortion of the material
adds a perturbation to $H_{\text{TLS}}$, which we denote by $H_1$. The
total Hamiltonian, 
$H=H_{\text{TLS}}+H_1$, enables us to describe the coupling between a
TLS and the phonon field. 
In the STM the variation of the off-diagonal elements of $H_{\text{TLS}}$ 
are neglected, so $H_1$ is
diagonal,\cite{philips:lowtemp,anderson:philmag,ZPhys.257.212.1972,RevModPhys.59.1,esquinazi:book}  
\begin{eqnarray}
H_1
 &=& \frac{1}{2}\left(\begin{array}{cc}
                \delta&0\\
                0&-\delta\end{array}\right)
                \,.\label{eqn_H_1_prime}
\end{eqnarray}
The perturbation $\delta$ is linear in the strain field, $S_{ij}$, 
\cite{RevModPhys.59.1,esquinazi:book} and in general may be written 
as $\delta\equiv 2\gamma_{ij} S_{ij}$. 
Here, as everywhere in this paper, we assume summation over repeated
indices.
The $3\times 3$ symmetric strain tensor is defined as 
$S_{ij}=\frac{1}{2}(\partial_iu_j+\partial_ju_i)$, with $u_i$, $i=1,2,3$, 
being the components of the displacement field. 
According to Eq.~(\ref{eqn_H_1_prime}), a static strain would just
slightly renormalize the eigenvalues of the Hamiltonian.
Since thermal and transport properties of an insulating solid are
determined by lattice vibrations we are interested in the effect of a
time-dependent strain entering the Hamiltonian $H_1$, 
describing the interaction of the TLS with a phonon field. 
In such a situation, the interaction produces transitions of the TLS 
between its eigenstates in the unperturbed state. 
These direct transitions contribute to the phonon mean free path and
in this way influence the heat conductance. 

The effect of the
time-dependent modulation of the spacing between the TLS energy levels
is very important for low-frequency
phonons.\cite{ZPhys.257.212.1972} The populations of the TLS levels
also change in time, but they lag behind the modulation of the
interlevel spacing. The resulting relaxation causes energy
dissipation and, in its turn, phonon damping. We do not consider this
effect here since the heat conductance is governed by thermal phonons
for which the resonant interaction is most important. 

In the STM the TLS interacts with 3D, transversally or longitudinally 
polarized, plane waves, which have all the components of the 
strain tensor proportional to the absolute value of the wave vector, $k$. 
The longitudinal wave will produce a compressional strain 
($S_{yz}=S_{zx}=S_{xy}=S_{zy}=S_{xz}=S_{yx}=0$) and the 
transversal wave will produce a shear strain 
($S_{xx}=S_{yy}=S_{zz}=0$). In this case, the expression for 
$\delta$ is always reduced to 
$\delta_\sigma = 2N_\sigma \gamma_\sigma k_\sigma$, where $\sigma\equiv t$ or 
$l$, denotes the transversal or longitudinal polarization of the wave,
$N_\sigma$ is the normalization constant of the phonon's displacement field and
$\gamma_\sigma$ is called the 
\textit{deformation potential parameter} or \textit{coupling constant}. 
Since $N_\sigma$ has dimensions of length, $N_\sigma k_\sigma$ is dimensionless 
and $\gamma_\sigma$ has dimensions of energy.
To calculate the transition rates of the TLS from one eigenstate to another, 
induced by the interaction with a phonon, we have to write $H_1$ 
in the basis that diagonalizes $H_{\text{TLS}}$: 
\begin{eqnarray}
H'_1
 &\equiv& O^TH_1 O = \frac{\delta}{2\epsilon}\left(\begin{array}{cc}
                             \Delta&-\Lambda\\
                            -\Lambda&-\Delta\end{array}\right)\,.
\label{eqn_H_1}
\end{eqnarray}

The off-diagonal terms in $H'_1$ determine the transition rates. 
Note that, since the excitation and de-excitation of the TLS cause
the absorption and emission of a phonon, respectively, $\delta$ is 
implicitly of 
the form $\delta\propto\bS_{\sigma\mathbf{k}}b_{\sigma\mathbf{k}}+h.c.$, where
$b_{\sigma\mathbf{k}}$ denotes a phonon annihilation operator. 
We omit this form here, but we shall use it explicitly in Section 
\ref{physical}.
To calculate the transition rates, let us denote the population of phonon 
modes by $n$ and the population of excited TLS states $f$. 
For example in thermal equilibrium, 
$n_{\bk\sigma}=[e^{\beta\hbar\omega_{\bk\sigma}}-1]^{-1}$ and 
$f_\epsilon = [e^{\beta\epsilon}+1]^{-1}$, where $\beta=1/(\kB T)$.
In these notations the TLS de-excitation
amplitude into a phonon of wave-vector $\bk$ and polarization $\sigma$, 
due to phonon-TLS interaction, is 
\begin{equation}
\langle n_{\bk\sigma}+1,\downarrow|H_1^\prime|n_{\bk\sigma}\uparrow\rangle = 
-\sqrt{\frac{\hbar k}{2V\rho c_\sigma}} \gamma_\sigma\frac{\Lambda}
{\epsilon} \sqrt{n_{\bk\sigma}+1} \,.\label{tranzH1}
\end{equation}
Using (\ref{tranzH1}), one obtains the contribution of a phonon with wave
vector $\mathbf{k}$ and polarization $\sigma$ to the TLS transition
probability due to phonon emission and absorption, respectively:
\begin{equation}
\Gamma_{\text{em}}(\epsilon,\bk,\sigma) = \gamma_\sigma^2 \frac{\pi k}{V\rho c_\sigma}\frac{\Lambda^2}{\epsilon^2}
(n_{\bk\sigma}+1) \delta(\hbar c_\sigma k-\epsilon) \,, \label{Gammaem}
\end{equation}
and 
\begin{equation}
\Gamma_{\text{abs}}(\epsilon,\bk,\sigma) = \gamma_\sigma^2 \frac{\pi
  k}{V\rho c_\sigma}\frac{\Lambda^2}{\epsilon^2} n_{\bk\sigma}
\delta(\hbar c_\sigma k-\epsilon) \,. \label{Gammaabs} 
\end{equation}
In the relaxation time approximation, summing 
$\Gamma_{\text{abs}}(\epsilon, \bk,\sigma)-\Gamma_{\text{em}}(\epsilon,\bk,\sigma)$ 
over all the phonon modes, we obtain the TLS  
relaxation time as\cite{ZPhys.257.212.1972,esquinazi:book} 
\begin{equation}
\tau_{\epsilon}^{-1} =
\left(\frac{\gamma^2_l}{c^5_l}+\frac{2\gamma^2_t}{c^5_t}\right)
\frac{\Lambda^2 \epsilon}{2\pi\rho \hbar^4}\cdot\coth\left(\frac{\beta
    \epsilon}{2}\right) \,. \label{tauTLS} 
\end{equation}
Similarly, we obtain the phonon relaxation time by summing over 
all the TLS states:\cite{ZPhys.257.212.1972,esquinazi:book}
\begin{equation}
\tau^{-1}_{\bk\sigma} = \frac{\pi\hbar\omega_{k,\sigma}}{\hbar\rho
  c^2_\sigma}\cdot\gamma_\sigma^2P_0\cdot \tanh\left(\frac{\beta
    \hbar\omega_{k,\sigma}}{2}\right) \,.  
\label{tauph}
\end{equation}

The low temperature mean free path of a phonon in the amorphous 
material, $c_\sigma \tau_{\bk\sigma}$, can be 
determined from the so-called unsaturated ultrasound attenuation, 
  i.~e., from the attenuation of an external acoustic wave of such small
  amplitude so that only a small fraction of TLSs are excited out of
  equilibrium. From the acoustic attenuation one can directly
  extract the product $\gamma_\sigma^2P_0$. 
Another way to determine $\gamma_\sigma^2P_0$ experimentally is by 
measuring the 
relative shift in the sound velocity:\cite{esquinazi:book,PhysRevB.17.2740} 
\begin{equation}
\frac{\Delta c_\sigma}{c_\sigma} = \frac{\gamma_\sigma^2P_0}{\rho c_\sigma^2}\ln\left(\frac{T}{T_0}\right)\,, \label{sound_vel_shift}
\end{equation}
where $T_0$ is a reference temperature at which $\Delta c_\sigma=0$. 
By these methods J. Black calculated the product $\gamma_\sigma^2P_0$. 
In fused silica,  for example, its values are within the interval
$(1.4\div4.6)\times 10^{7}$ J/m$^3$  for longitudinal
phonons and $(0.63-0.89)\times 10^{7}$ J/m$^3$ for 
transversal phonons, see Ref.~\onlinecite{PhysRevB.17.2740} and 
references therein.

The values obtained for $\gamma_\sigma^2P_0$ by these two independent 
methods can be combined to calculate the heat conductivity as
\begin{equation}
\kappa(T) = \frac{\rho \kB^3}{6\pi\hbar^2}\left(\frac{c_l}{\gamma_l^2 P_0} 
+ \frac{2c_t}{\gamma_t^2 P_0}\right)T^2 \,, \label{kappa1}
\end{equation}
which can be measured by yet another experiment.\cite{PhysRevB.8.2896} 

Although sometimes there are pronounced  differences between the values for 
$\gamma_\sigma^2P_0$ obtained in different experiments (we are not 
concerned here with classification of results), it seems that always 
$\gamma_l>\gamma_t$. 
To the best of our knowledge, this aspect is not explained in the literature.
Moreover, the 
simplified expression for the perturbation term $\delta$ in the interaction 
Hamiltonian does not allow us to calculate the interaction of the 
TLS with an arbitrary strain field. Since, as mentioned above, the elastic 
modes in mesoscopic systems have rather complicated displacement and strain 
fields, we need a more general interaction Hamiltonian which 
should incorporate the microscopic symmetry of the material around 
the TLS and the orientation of the TLS with respect to the strain field.
In the next section we will build such an interaction Hamiltonian and, 
after we apply it to 3D bulk systems, we will show how we can extract 
information about its parameters from experiments. 
The relationship $\gamma_l>\gamma_t\ge 0$ 
is a natural result of our model.
We will provide the results for the interaction between TLSs and Lamb modes
in ultra-thin membranes elsewhere.

%
\section{The general TLS-phonon interaction Hamiltonian} 

To generalize the TLS Hamiltonian we shall use the full expression 
for $\delta$ of Eq. (\ref{eqn_H_1_prime}),
\begin{equation}
\delta = 2\gamma_{ij}S_{ij} \equiv 2[\gamma]:[S] \,, \label{eqn_H_1gen}
\end{equation}
where $\gamma_{ij}$ are the components of the 
$3\times 3$ tensor $[\gamma]$ and ``$:$'' is the symbol for the dyadic
product.  
Let us now find the general properties of $[\gamma]$. 

As follows from Eq. (\ref{eqn_H_1gen}), only the symmetric part of the
tensor $[\gamma]$ has a physical meaning. Indeed, the dyadic product
between a symmetric and an antisymmetric tensor is zero, so the 
antisymmetric part of $[\gamma]$, even if existent, would not influence the results.
We therefore assume $[\gamma]$ to be symmetric.
Since all the tensors we have in our model are symmetric, it is 
more convenient to work using the \textit{abbreviated subscript notation}, 
as described for example in Ref.~\onlinecite{Auld:book}. We will explain this 
method briefly. 

Let us assume that $[A]$ is a symmetric $3\times 3$ tensor 
(i.e. $A_{ij}=A_{ji}$). From its 
9 elements, only 6 are independent. To get rid of the redundant 
3 elements and also to make the tensors manipulation easier, we can write 
$[A]$ in the form of a six component vector, $\bA$, in the 
following 2 ways:\cite{Auld:book}
\begin{eqnarray}
&& A_1 \equiv A_{xx},\ A_2 \equiv A_{yy},\ A_3 \equiv A_{zz}, 
\nonumber \\
&& A_4 \equiv A_{yz},\ A_5 \equiv A_{zx},\ A_6 \equiv A_{xy} 
\label{def_abbr_subs1}
\end{eqnarray}
or 
\begin{eqnarray}
&& A_1 \equiv A_{xx},\ A_2 \equiv A_{yy},\ A_3 \equiv A_{zz}, 
\nonumber \\
&& A_4 \equiv 2A_{yz},\ A_5 \equiv 2A_{zx},\ A_6 \equiv 2A_{xy} 
\label{def_abbr_subs2}
\end{eqnarray}
Then the dyadic product of two symmetric tensors, $[A]:[B]$, 
may be written as $[A]:[B]=\bA^T\cdot\bB$ if one of the tensors is 
written in the form (\ref{def_abbr_subs1}) and the other one in the form 
(\ref{def_abbr_subs2}). 
Applying this way of writing to equation (\ref{eqn_H_1gen}), we 
define
$\bgamma\equiv(\gamma_{xx},\gamma_{yy},\gamma_{zz},\gamma_{yz},\gamma_{zx},\gamma_{xy})^T$
and $\bS\equiv(S_{xx},S_{yy},S_{zz},2S_{yz},2S_{zx},2S_{xy},)$, so
that $\delta=2\mbox{\boldmath$\gamma$}^T\cdot\bS$.  

As in Ref.~\onlinecite{Auld:book}, the subscripts in the abbreviated subscript 
notations will be denoted by capital letters, $I,J,K,\ldots$, which run 
from 1 to 6. 
%

\subsection{The properties of the deformation potential tensor}

The vector $\mbox{\boldmath$\gamma$}$ should characterize the TLS and 
its deformability 
in the presence of a strain field. As explained before, the TLS is 
imagined as a particle or a group of particles that tunnels from one 
potential well to another. This tunneling may happen as a translation 
between the wells, or as a rotation.\cite{galperinetal:AdvinP} In either 
case, there is a direction 
associated to the TLS, which we call $\hat\bt$, for example the direction 
defined by the two potential wells or the axis of rotation. 
One can expect that the orientation of the TLS (i.~e., $\hat\bt$) 
relative to the phonon's propagation direction and polarization has 
an effect on the interaction strength. The three components of $\hat{\bt}$
are the only co(ntra)variant quantities that describe the TLS, from a 
very general point of view, i.e. without building a microscopic model of
the TLS. 
With these 
quantities we can construct the simplest symmetric $3\times 3$ tensor 
\begin{eqnarray*}
[T] &\equiv& \left(\begin{array}{ccc} t_x^2 & t_xt_y & t_xt_z \\ t_xt_y & t_y^2 & t_yt_z \\ t_xt_z & t_yt_z & t_x^2 \end{array}\right) = \hot\cdot\hot^T ,
\end{eqnarray*}
and a general one, $[\bgamma]=[R]:[T]$ (i.e. $\bgamma_{kl}=R_{ijkl}T_{ij}$), 
with $R_{ijkl}=R_{ijlk}$ for any $k$ and $l$. We shall also choose 
$R_{ijkl}=R_{jikl}$, since the summation $R_{ijkl}T_{ij}$ allows us to 
use this simplification. In abbreviated subscript notations, 
$\bT=(t_x^2,t_y^2,t_x^2,2t_yt_z,2t_zt_x,2t_xt_y)^T$, 
and $R_{ijkl}$ becomes $R_{IJ}$ in an obvious way. 
Then we can write $\bgamma$ as 
\begin{equation}
\mbox{\boldmath$\gamma$} \equiv [R]^T\cdot\bT \,. \label{split_gamma}
\end{equation}
Since $\bT$ characterizes the orientation of the TLS, the relevant
deformation potential parameters are contained in $[R]$. To make an 
analogy, the tensor $[R]$ is similar to the tensor of 
\textit{elastic stiffness constants} from elasticity theory. 
Still, the matrix $[R]$ cannot be taken arbitrary. Like the 
elastic stiffness constants, the matrix $[R]$ is determined by the 
\textit{local} symmetry properties of the atomic lattice, around the TLS. 
We will deduce here the properties of $[R]$. 

We start by noting that the product $h_1\equiv\bT^T\cdot[R]\cdot\bS$ 
is a scalar, so it should be invariant under any rotation of coordinates. 
Moreover, since $[R]$ reflects the microscopic symmetry of 
the lattice around the TLS, we 
shall do a sequence of transformations to obtain the properties of $[R]$. 
For the moment, let us assume that the lattice is simple cubic: 
\begin{itemize}
\item Let us choose $S_1=S_2=S_3=S_5=S_6=0$, $S_4\ne0$ and $\hat\bt=\hat\bx$. 
Then we 
rotate the coordinate system through $\pi$ about the $z$ axis. Under this 
rotation $[R]$ is invariant, whereas $\bT=(t_x^2,0,0,0,0,0)^T$ transforms 
into $\bT'=\bT$ and $\bS=(0,0,0,S_4,0,0)^T$ transforms into $\bS'=-\bS$. Then, 
$h_1 = \bT\cdot[R]\cdot\bS=R_{14}t_x^2S_4=\bT'\cdot[R]\cdot\bS'=-R_{14}t_x^2S_4$ 
implies that $R_{14}=0$. 
Performing similar rotations about the other axes of the coordinate system
with appropriately chosen $\bT$ and $\bS$, we can show that
$R_{IJ}=0$ for any $I=1,2,3$ and $J=4,5,6$.
\item Taking $\bS=(S_1\ne0,0,0,0,0,0)^T$ and $\bT=(0,t_y^2,t_z^2,2t_yt_z,0,0)^T$
(so $t_x=0$ and $t_y,t_z\ne0$) and rotating the coordinate system through 
$\pi$ about the $z$ axis, we transform 
$\bS$ into $\bS'=\bS$ and $\bT$ into 
$\bT^\prime=(0,t_y^2,t_z^2,-2t_yt_z,0,0)^T$. From the product 
$h_1^\prime=S_1(R_{21}t_y^2+R_{31}t_z^2-R_{41}2t_yt_z)
=S_1(R_{21}t_y^2+R_{31}t_z^2+R_{41}2t_yt_z)=h_1$, we deduce that $R_{41}=0$. 

Again, performing similar rotations about the other coordinate axes
we can show that $R_{JI}=0$ for any $I=1,2,3$ and $J=4,5,6$. 

Up to now we proved that the matrix $[R]$ is block-diagonal. 
Let us see if we can simplify it further. 
\item Assume that $t_x=0$, whereas $t_y$ and $t_z$ are different from 
zero. If, moreover, from the components of the strain vector only $S_5$ is 
different from zero and we rotate the coordinates through 
$\pi$ about the $x$ axis, then 
$\bT=(0,t_y^2,t_z^2,2t_yt_z,0,0)^T\to\bT=(0,t_y^2,t_z^2,2t_yt_z,0,0)^T$ and 
$\bS=(0,0,0,0,S_5,0)\to\bS'=(0,0,0,0,-S_5,0)$ 
Calculating the products $\bT^T\cdot[R]\cdot\bS$ and $(\bT')^T\cdot[R]\cdot\bS'$ we 
find $R_{45}=0$. 

By similar arguments we conclude that 
$R_{45}=R_{46}=R_{56}=R_{54}=R_{64}=R_{65}=0$. 
\item Due to the cubic symmetry, $h_1$ and $H_1$ should not 
depend on the notation of axes. Therefore $R_{11}=R_{22}=R_{33}$, 
$R_{12}=R_{13}=R_{23}$, $R_{21}=R_{31}=R_{32}$, and $R_{44}=R_{55}=R_{66}$. 
\end{itemize}
These are all the constraints that we can impose on $[R]$ if the lattice 
around the TLS has cubic symmetry. Now let us make one more simplification 
and assume the system is isotropic and find a relationship between the 
parameters $R_{11}$, $R_{12}$, $R_{21}$, and $R_{44}$. 

Again from elasticity theory we know that both $\bT$ and $\bS$ 
transform under a rotation of coordinates like 
$\bT'=[N]\cdot\bT$ and $\bS'=[N]\cdot\bS$, where the matrix $[N]$ is defined 
for example at page 75 of Ref.~\onlinecite{Auld:book}. Since at a rotation 
of coordinates only $\bT$ and $\bS$ change, and not the deformation 
potential tensor (the interaction Hamiltonian 
should look the same in any coordinate system), 
$h_1=\bT^T\cdot[R]\cdot\bS=\bT^T\cdot[N]^T\cdot[R]\cdot[N]\cdot\bS$ for 
any $\bT$ and $\bS$. This implies $[R]=[N]^T\cdot[R]\cdot[N]$. 
Taking arbitrary rotations about each of the coordinate axes, we obtain 
the final conditions: $R_{12}=R_{21}=R_{13}=R_{31}=R_{32}=R_{23}$ and 
$R_{11}-2R_{44}=R_{12}$. 

By denoting $R_{11}\equiv\tilde\gamma$, $R_{12}/\tilde\gamma\equiv\zeta$, and 
$R_{44}/\tilde\gamma\equiv\xi$, we arrive at 
\begin{eqnarray}
[R] &=& \tilde\gamma\cdot \left(\begin{array}{cccccc}
                1&\zeta&\zeta&0&0&0\\
                \zeta&1&\zeta&0&0&0\\
                \zeta&\zeta&1&0&0&0\\
                0&0&0&\xi&0&0\\
                0&0&0&0&\xi&0\\
                0&0&0&0&0&\xi\end{array}\right) 
\equiv \tilde\gamma\cdot[r]\,,\label{eqn_R}
\end{eqnarray}
with $\zeta+2\xi=1$. 

The form of $[R]$ is very general, yet, the parameters $\tilde\gamma$, 
$\xi$ and implicitly $\zeta$ may vary from one type of TLS to another. 
For example, for the two types of TLSs mentioned in the beginning of 
this subsection--translational and rotational (see
Ref.~\onlinecite{galperinetal:AdvinP}  
and references therein)-- the parameters of $[R]$ may be different.

\subsection{Physical results} \label{physical}

We can now assemble back the expression for $\delta$: 
\begin{equation}
\delta = 2\tilde\gamma\bT^T\cdot[r]\cdot\bS \,. \label{delta_do2}
\end{equation}
For calculations of physical quantities, we have to write $H_1$ in 
second quantization.\cite{submitted.quantization} 
First, we introduce the excitation and de-excitation 
operators for the TLS, 
\begin{equation}
a^\dagger=\left(\begin{array}{cc}0&1\\0&0\end{array}\right),\quad 
a=\left(\begin{array}{cc}0&0\\1&0\end{array}\right) \,.
\nonumber
\end{equation} 
These matrix operators obey Fermionic commutation relations and satisfy 
the conditions: $\sigma_z=(2a^\dagger-1)$ and $\sigma_x=(a^\dagger+a)$. 
The phonon creation and annihilation operators are denoted by 
$b^\dagger_\mu$ and $b_\mu$, respectively. Here we use $\mu$ to 
denote general phonon modes. For 3D plane waves, $\mu\equiv(\bk\sigma)$. 

With these definitions, in the basis that diagonalizes $H_{\text{TLS}}$, the 
interaction Hamiltonian reads: 
\begin{eqnarray}
\tilde{H}_1
 &=&\frac{\tilde\gamma\Delta}{\epsilon}\mathbf{T}^T\cdot[r]\cdot
         \sum_{\mu}\left[\mathbf{S}_\mu b_\mu 
                      +\mathbf{S}^\star_\mu b^\dagger_\mu
\right](2a^\dagger a-1)
    \nonumber\\
 && \hspace{-2.5mm}
  -\frac{\tilde\gamma\Lambda}{\epsilon}\mathbf{T}^T\cdot[r]\cdot
         \sum_\mu\left[\mathbf{S}_\mu b_\mu 
                      +\mathbf{S}^\star_\mu b^\dagger_\mu 
\right](a^\dagger+a)
    \,,\label{eqn_H_1_tilde}
\end{eqnarray}
where we used the notation $\tilde{H}_1$ to distinguish this form from 
the one in Eq.\ (\ref{eqn_H_1}). By $\mathbf{S}_\mu$ we denote the 
strain produced by the phonon field $\mu$ at the position of the TLS. 

In first order perturbation theory, the matrix element for 
the absorption of a phonon $(\bk\sigma)$ by an unexcited TLS is 
\begin{equation}
\bracket{n_{\bk\sigma},\uparrow}{\tilde{H}_1|n_{\bk\sigma}+1,\downarrow}
 = -\frac{\tilde\gamma\Lambda}{\epsilon}\sqrt{\frac{\hbar n_{\bk\sigma}}{2V\rho
\omega_{\bk\sigma}}}
     \mathbf{T}^T\cdot[r]\cdot\mathbf{S}_{\bk\sigma}\,. 
\label{eqn_matrix_element}
\end{equation}
Applying Fermi's golden rule, we obtain the transition
probability for this process. Assuming that in an amorphous solid the
TLS directions 
are uniformly distributed, we sum the contributions of the TLSs
averaging over their 
directions and we obtain an average transition rate, 
\begin{equation}
\overline{\Gamma}_{\ket{n_{\bk\sigma},\uparrow},\ket{n_{\bk\sigma}+1,\downarrow}}   
 = C_\sigma\tilde\gamma^2\frac{n_{\bk\sigma}\pi k}{V\rho c_\sigma}
 \cdot\frac{\Lambda^2}{\epsilon^2} 
\delta(\hbar\omega_{\bk\sigma}-\epsilon)\,,\label{eqn_Gamma_bar}
\end{equation}
where $C_\sigma$ is a constant that depends on the polarization of the phonon. 
Similarly, the emission rate is 
\begin{equation}
\overline{\Gamma}_{\ket{n_{\bk\sigma}+1,\uparrow},\ket{n_{\bk\sigma},\downarrow}}
 = C_\sigma\tilde\gamma^2\frac{(n_{\bk\sigma}+1)\pi k}{V\rho c_\sigma}
 \cdot\frac{\Lambda^2}{\epsilon^2}\delta(\hbar\omega_{\bk\sigma}-\epsilon)\,.\label{eqn_Gamma_bar_em}
\end{equation}
Equations (\ref{eqn_Gamma_bar}) and (\ref{eqn_Gamma_bar_em}) are similar 
to equations (\ref{Gammaabs}) and (\ref{Gammaem}) respectively, with 
$\gamma_t^2$ replaced by $C_t\tilde\gamma^2$ and $\gamma_l^2$ replaced by 
$C_l\tilde\gamma^2$. The constants $C_t$ and $C_l$ are 
\begin{subequations}
\begin{eqnarray}
C_{l}
 &=& \frac{1}{15}(15-40\xi+32\xi^2)\label{eqn_N_l}\\
C_{t}
 &=& \frac{4}{15}\xi^2\,.\label{eqn_N_t}
\end{eqnarray}
\end{subequations}
Even though we are not able to make any statement about the range 
in which $\xi$ takes value, we can still make the interesting prediction: 
\begin{equation}
C_l>C_t \ge 0\ {\rm for\ any}\ \xi, \label{Nprediction}
\end{equation}
in agreement with the experimental data. 

The TLS relaxation time and phonon absorption time can still be calculated 
by Eqs. (\ref{tauTLS}) and (\ref{tauph}), with $\gamma_t^2$ and 
$\gamma_l^2$ replaced by $C_t\tilde\gamma^2$ and $C_l\tilde\gamma^2$. 
Therefore, $C_t\tilde\gamma^2$ and $C_l\tilde\gamma^2$ can be calculated 
from unsaturated ultrasonic attenuation or sound velocity shift experiments. 
Once the values $C_t\tilde\gamma^2$ and $C_l\tilde\gamma^2$ are obtained, 
the ratio 
\begin{equation}
\frac{4C_{l}}{C_{t}} = \frac{15}{\xi^2}-\frac{40}{\xi}+32 \label{eqxi} 
\end{equation}
gives us the value of $\xi$, which further enables us to calculate 
$\zeta$. If the value of $\tilde\gamma$ can be extracted from phonon echo 
experiments,\cite{PhysRevLett.37.852} then all the elements of the 
displacement potential tensor $[R]$ are known. 

As a numerical example, we take from Ref.~\onlinecite{PhysRevB.17.2740}
the two sets of values for $P_0\gamma_l^2$ and $P_0\gamma_t^2$ that we 
used before. For the first one,\cite{PhysRevB.14.1660} where 
$P_0\gamma_l^2=1.4\times10^{-5}$ J/m$^3$ and 
$P_0\gamma_t^2=0.63\times 10^{-5}$ J/m$^3$, the two solutions for $\xi$ are 
$\xi_1=0.55$ and $\xi_2=1.2$. Using the formula 
$P_0\tilde\gamma^2=P_0\gamma_t^2/C_t(\xi)$ (see Eq. \ref{eqn_N_t}) 
we obtain $(P_0\tilde\gamma^2)_1=7.8\times 10^7$ and 
$(P_0\tilde\gamma^2)_2=1.7\times 10^7$. For the second set of 
values\cite{PhysRevLett.37.1248}--$P_0\gamma_l^2=2.0\times 10^{-5}$
J/m$^3$ and $P_0\gamma_t^2=0.89\times 10^{-5}$ J/m$^3$--we get 
$\xi_1=0.55$ and $\xi_2=1.2$, 
which correspond to $(P_0\tilde\gamma^2)_1=11\times 10^7$ and 
$(P_0\tilde\gamma^2)_2=2.4\times 10^7$.
Note that although $P_0\tilde\gamma^2$ (or $P_0\tilde\gamma^2_\sigma$) 
changes significantly from one experiment to another, the ratio 
$P_0\gamma_l^2/P_0\gamma_t^2$ does not change much, which leads to close 
values for the parameter $\xi$. 
From these measurements only, we cannot make the difference between 
$\xi_1$ and $\xi_2$.

\section{Conclusions}

To describe the interaction of a TLS with an arbitrary strain field, 
we introduced a generalization of the standard TLS-plane wave interaction 
Hamiltonian. Such a generalization is useful in the description of 
mesoscopic systems, where the phonon modes are not anymore the simple, 
transversally and longitudinally polarized
plane waves, but have more complicated displacement fields and dispersion 
relations. We 
used the symmetry properties of the system to deduce the properties of 
our interaction Hamiltonian, $\tilde H_1$. 

We showed that if the TLS is in an isotropic medium, then 
$\tilde H_1$ depends on four free parameters: two from the 
orientation of the TLS, $\hat\bt$, 
and another two, denoted $\tilde\gamma$ and $\xi$ (\ref{eqn_R}), which 
describe the coupling of the oriented TLS with the strain field. 
Since, in an amorphous solid, $\hat\bt$ has random, uniformly distributed 
directions, the first two parameters are averaged 
and the effective form of the Hamiltonian has only two free parameters, 
$\tilde\gamma$ and $\xi$, like the Hamiltonian of the standard tunneling 
model. From sound absorption, sound velocity change, heat conductivity and 
phonon-echo experiments one can calculate these two parameters, like we 
did in the end of the previous section. 

The other way around, if for example the distribution of the TLS orientations 
is not uniform, knowing the values of the parameters 
$\tilde\gamma$ and $\xi$ one can eventually reconstruct the
distribution.

Different types of TLSs--translational and rotational ones--may have 
different deformation potential parameters. For one type of TLSs 
we proved that $\gamma_l>\gamma_t$, 
which is confirmed by all the experimental results we know about. 

\acknowledgments 
Discussions with J. P. Pekola, I. Maasilta, and V. Vinokur are
gratefully  acknowledged. 
This work was partly supported  by the U. S.
Department of  Energy Office of Science through contract No.
W-31-109-ENG-38 and by the NATO grant 
PDD(CP)-(CPB.EAP.RIG 982080).


\end{document}